\documentclass[aps,prb,twocolumn,showpacs,superscriptaddress,groupedaddress]{revtex4}

\usepackage{natbib}
\usepackage{graphicx} 
\usepackage{amsmath} 
\usepackage{graphicx}
\usepackage{setspace}

\usepackage[british]{babel}

\newcommand{\eq}[1]{Eq.~(\ref{eq:#1})}

\begin{document}
 
\title{Adiabatic quantum pumping of chiral Majorana fermions}

\author{M. Alos-Palop} 
\affiliation{Delft University of Technology, Kavli Institute of
  Nanoscience, Department of Quantum Nanoscience, Lorentzweg 1, 2628
  CJ Delft, The Netherlands.}
 
\author{Rakesh P. Tiwari} 
\affiliation{Department of Physics, University of Basel,
  Klingelbergstrasse 82, CH-4056 Basel, Switzerland}

\author{M. Blaauboer} 
\affiliation{Delft University of Technology, Kavli Institute of
  Nanoscience, Department of Quantum Nanoscience, Lorentzweg 1, 2628
  CJ Delft, The Netherlands.}

\date{\today}

\begin{abstract}

We investigate adiabatic quantum pumping of chiral Majorana states in
a system composed of two Mach--Zehnder type interferometers coupled
via a quantum point contact. The pumped current is generated by
periodic modulation of the phases accumulated by traveling
around each interferometer. Using scattering matrix formalism we show
that the pumped current reveals a definite signature of the chiral
nature of the Majorana states involved in transport in this geometry. 
Furthermore, by tuning the coupling between the two
interferometers the pump can operate in a regime where finite pumped
current and zero two-terminal conductance is expected.

\end{abstract}


\maketitle

\textit{Introduction}.\ Recently, a great amount of attention has
been paid to the possibility of realizing Majorana quasiparticles in
condensed matter systems~\cite{Alicea2012, Leijnse2012,
  Beenakker2011}.  Majorana-like excitations have been predicted to
exist in the $\nu~=~5/2$ quantum Hall state~\cite{Moore1991,
  Stern2001}, p--wave superconductors~\cite{Ivanov2001},
semiconductor--superconductor interfaces~\cite{Sau2010a, Sau2010b,
  Alicea2010} and on the surface of topological insulators
~\cite{Hasan2010, Qi2011, Fu2008, Tiwari2012}.  Zero-bias conductance
anomalies~\cite{zba1,zba2,Law2009,zba3} associated with localized
Majorana excitations have been measured recently~\cite{Mourik2012,
  Das2012, Deng2012}. Measurements of unconventional Josephson effects
associated with these excitations have also been
reported~\cite{Rokhinson2012, Williams2012}. In addition, unique
signatures of \textit{chiral} Majorana fermions have been predicted in
Mach--Zehnder~\cite{Fu2009, Akhmerov2009} and Hanbury
Brown--Twiss~\cite{Strubi2011} type interferometers through
conductance and noise measurements. In this article, we propose and
analyze an adiabatic Majorana quantum pump which can provide a conclusive
evidence of the chiral nature of the Majorana modes.

\begin{figure}
  \centering
  \includegraphics[width=.8\columnwidth]{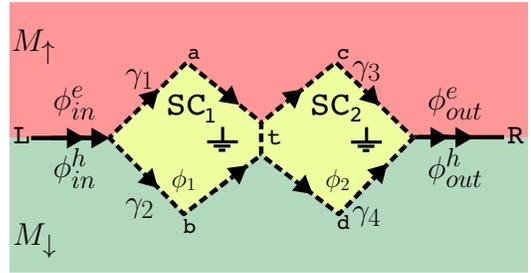}
  \caption{(color online). Schematic of the Mach--Zehnder
    interferometer studied in this paper. Two superconducting islands,
    SC$_1$ and SC$_2$, are connected via a point contact (t). In the
    left ({\it in}) and right ({\it out}) lead, two chiral Dirac
    fermion modes, $\phi^e$ and $\phi^h$, propagate. The entire set-up
    is placed on top of a 3D topological insulator. Majorana fermions
    $\gamma_l$ ($l=$ 1--4) are the mediating states in the central
    interferometer region. See the text for more details.}
  \label{fig:MZIschematic}
\end{figure}

Adiabatic pumping is a transport mechanism in meso- and nanoscale
devices by which a finite dc current is generated in the absence of an
applied bias by low-frequency periodic modulations of at least two
system parameters (typically gate voltages or magnetic
fields)~\cite{Buttiker1994, Brouwer1998}. In order for electrical
transport to be adiabatic, the period of the oscillatory driving
signals has to be much longer than the dwell time $\tau_{\rm dwell}$
of the electrons in the system, $ T = 2 \pi \omega^{-1} \gg \tau_{\rm
  dwell} $. Adiabatic {\it quantum} pumping~\cite{Spivak1995} refers
to pumping in open systems in which quantum-mechanical interference of
electron waves occurs. Recently, adiabatic topological pumping in a
spin-orbit coupled semiconductor nanowire in proximity to a s-wave
superconductor and subjected to a Zeeman field was
studied~\cite{Gibertini2013}. In this study we consider an adiabatic
quantum pump where the carriers responsible for transport are
\textit{chiral} Majorana fermions.  A schematic of the proposed device
is shown in Fig.~\ref{fig:MZIschematic}. The pump consists of two
superconducting islands supporting chiral Majorana edge states coupled
via a quantum point contact. While the conductance in this system can
be used to signal whether an unpaired Majorana bound state exists in
the superconducting region or not (as was predicted in
Refs.~\onlinecite{Fu2009, Akhmerov2009}), it does not contain
information about the chiral nature of the carriers. We show that, in
contrast, the pumped current in this system exhibits definite and
measurable signatures of the chiral nature of quantum transport.
Furthermore, charge neutrality of the Majorana modes (limiting
interactions with the environment) and the adiabatic operation of the
pump makes this system attractive for studying quantum interference
effects with Majorana modes in the presence of, in principle,
negligible dephasing.

\textit{Majorana quantum pump}.\ The superconducting and magnetic
correlations on the surface of a topological insulator (as shown in
Fig.~\ref{fig:MZIschematic}, with $\hat{z}$ being the unit vector in
the out-of-plane direction), as well as the gapless states at the 
interfaces between them can be described by the Dirac--Bogoliubov--de
Gennes (DBdG) Hamiltonian $H=\Psi^{\dagger}\mathcal{H} \Psi/2$. In the
Nambu basis where $\Psi=(u_{\uparrow}, u_{\downarrow},
v_{\downarrow},-v_{\uparrow})^T$~\cite{Fu2009}, the Hamiltonian
$\mathcal{H}$ is given by
\begin{eqnarray}
\mathcal{H}=&-&i\hbar v_F \tau_{z}\otimes\hat{z}\cdot\vec{\bf
  \sigma}\times{\bf \nabla} -\mu \tau_z\otimes\sigma_0 \nonumber
\\ &+&(\Delta\tau_{+}\otimes\sigma_0
+\Delta^{\ast}\tau_{-}\otimes\sigma_0) + M\tau_0\otimes\sigma_z.
\label{eq:DBdG}
\end{eqnarray}
Here $\vec{\sigma}$ and $\vec{\tau}$ represent vectors of Pauli
matrices in spin space and particle--hole Nambu space respectively.
Similarly, $\sigma_0$ and $\tau_0$ represent 2$\times$2 identity
matrices in spin and Nambu space, and $\tau^{\pm}=(\tau_{x}\pm
i\tau_y)/2$. The first two terms in $\mathcal{H}$ describe the free
surface states of the three-dimensional topological insulator with
$v_F$ the Fermi velocity and $\mu$ the chemical potential.  We choose
the coordinate system such that this surface is parallel to the $x-y$
plane.  The first term in the second line in $\mathcal{H}$ describes
the superconducting proximity effect due to a s-wave superconductor.
The magnetizations $M_{\uparrow}=(0,0, M)$ and
$M_{\downarrow}=(0,0,-M)$ of the two ferromagnetic insulators (as
shown in Fig.~\ref{fig:MZIschematic}) describe the effect of Zeeman
splitting as expressed by the last term in
Eq.~(\ref{eq:DBdG}). $\Delta$ and $M$ are assumed to be spatially
uniform.  The excitation spectrum is gapped in both the
superconducting and the magnetic regions. In the superconducting
region the excitation spectrum is $E_{\bf k}^{\mathcal{S}}=\sqrt{(\pm
  v_{F} |{\bf k}|-\mu)^2+|\Delta|^2}$. In the magnetic region it is
$E_{\bf k}^{\mathcal{M}}=\sqrt{v_{F}^2 |{\bf k}|^2+M^2} \pm \mu$
(which is gapped if $M>\mu$).  Solutions of Eq.~(\ref{eq:DBdG}) also
include the subgap chiral Majorana branch localized near the
superconductor--ferromagnet interface with group velocity
$v_m=v_F\sqrt{1-\mu^2/M^2}/(1+\mu^2/|\Delta|^2)$ ~\cite{Fu2009}.  The
amplitudes of these chiral Majorana modes are denoted by $\gamma_l$,
$l\in\{1,4\}$ in Fig.~\ref{fig:MZIschematic}. The interface between
regions with opposite signs of magnetizations supports two chiral
Dirac fermion modes. One is the electron mode with amplitude $\phi^e$
and the other the hole mode with amplitude $\phi^{h}$.  Within the
Landauer--B\"{u}ttiker scattering matrix formalism we can relate the
two incoming modes $\phi^e_{in}$ and $\phi^h_{in}$ with two outgoing
modes $\gamma_1$ and $\gamma_2$ at the left tri-junction using
$(\gamma_1,\gamma_2)^T=S(E)(\phi^e_{in},\phi^h_{in})^T$. Particle-hole
symmetry [$S(E)=S^{\ast}(-E)\tau_x$] along with unitarity
[$(S^{\dagger})^{-1}=S$] allows us to choose at $E=0$
\begin{equation}
S = \frac{1}{\sqrt{2}} \left( \begin{array}{cc}
      1 & 1 \\ i & - i \\
\end{array} \right).
\label{eq:0energys}
\end{equation}
Similarly we can relate the chiral Majorana modes $\gamma_3$ and
$\gamma_4$ to outgoing electron and hole modes at the right
tri-junction. In the following it is assumed that $S(E)$ is well
described by its zero energy limit, which is appropriate for small
energies $E\ll(v_m/v_F)\text{min}(|\Delta|,M)$ and junctions with
mirror symmetry~\cite{Fu2009}.

The Majorana modes $\gamma_1$ and $\gamma_2$ are coupled to Majorana
modes $\gamma_3$ and $\gamma_4$ via the Josephson junction between the
two superconductors (denoted as SC$_1$ and SC$_2$ in
Fig.~\ref{fig:MZIschematic}). The junction acts as a quantum point
contact (QPC) for the Majorana modes and can be characterized by a
2$\times$2 scattering matrix,
$(\gamma_3,\gamma_4)^T=S_{QPC}(\gamma_1,\gamma_2)^T$~\cite{Fu2009}
where
\begin{equation}
 S_{QPC}=\left(\begin{array}{cc} r_1 & t_2 \\
           t_1 & r_2 
           \end{array}
\right).
\end{equation}
Here $|t_{1}|^2 = 1 - |r_1|^2$ and $|t_{2}|^2 = 1 - |r_{2}|^2$. The
properties of this QPC can be tuned by changing the phase difference
$\phi_1 - \phi_2$ of the Josephson junction (as shown in
Fig.~\ref{fig:MZIschematic}) or by altering its shape. As explained in
Ref.~\onlinecite{Fu2009}, this Josephson junction describes
superconductors weakly coupled by single electron tunneling at a
point. Particle--hole symmetry and unitarity imply that $r_j$ and
$t_j$ are real coefficients. Below we assume a symmetric Josephson
junction and set $r_1=r_2=r$, and $t_1=-t_2=t$. The incoming electrons
and holes can be related to the outgoing electrons and holes by the
full scattering matrix of the system
$(\phi^e_{out},\phi^h_{out})^{T}=S_{RL}(\phi^e_{in},\phi^h_{in})^{T}$.
The scattering matrix $S_{RL}$ can be decomposed into
$S_{RL}=S^{\dagger}S_2S_{QPC}S_{1}S$, where
\begin{equation}
S_1=\left(\begin{array}{cc} e^{i \beta_a} & 0
  \\ 0 & e^{i\beta_b}
          \end{array}
\right), 
\;\;
S_2=\left(\begin{array}{cc} e^{i \beta_c} & 0
  \\ 0 & e^{i\beta_d}
          \end{array}
\right),
\end{equation}
denote the contribution from the phase shifts $\beta_k$
($k\in\{a,b,c,d\}$) picked up by the Majorana modes by traversing the
$k^{\text{th}}$ arm of the interferometer. The relative phase shifts
$\beta_a - \beta_b\equiv \tilde{\theta}_1 = \pi n_{\nu_1} + \pi + E
\delta\tau_1/\hbar$ includes a contribution of $\pi$ for every vortex
enclosed, a Berry phase of $\pi$ for spin-1/2 particles and the
dynamical phase. Similarly, $\beta_c - \beta_d \equiv
\tilde{\theta}_2=\pi n_{\nu_2} + \pi + E \delta\tau_2/\hbar$.  Here
$n_{\nu_1}$ and $n_{\nu_2}$ denote the number of vortices in $SC_1$
and $SC_2$, $\delta\tau_1=L_a/(v_m)_a-L_b/(v_m)_b$,
$\delta\tau_2=L_c/(v_m)_c-L_d/(v_m)_d$ where $L_{k}$ and $(v_m)_k$ are
the length and the velocity of the chiral Majorana mode in the
$k^{\text{th}}$ arm of the interferometer.  We then obtain
\begin{eqnarray}
S_{RL} & = &e^{i(\beta_b + \beta_d)} \left[ \begin{array}{cc} \eta_1^+
    r - i \eta_2^+ t & -\eta_1^- r -i \eta_2^- t \\ -\eta_1^- r +i
    \eta_2^- t & \eta_1^+r + i \eta_2^+ t \\
\end{array} \right] ,
\label{eq:MZscatteringmatrix}
\end{eqnarray}
where $\eta_1^\pm = (1 \pm e^{i(\tilde{\theta}_1 +
  \tilde{\theta}_2)})/2$, $\eta_2^\pm = (e^{i \tilde{\theta}_1} \pm
e^{i \tilde{\theta}_2})/2$ and the $(2,1)$-element of $S_{RL}$
indicates conversion of an incoming electron in the left lead to an
outgoing hole in the right lead.

\textit{Adiabatic quantum pumping\/}. In our device the adiabatically
pumped current through the Mach--Zehnder interferometer is driven by
periodic modulation of the phases $\tilde{\theta}_1$ and
$\tilde{\theta}_2$ as $\tilde{\theta}_1 (t) = \theta_{1} + \delta
\theta_1 \cos( \omega t)$ and $\tilde{\theta}_2 (t) = \theta_{2} +
\delta \theta_2 \cos( \omega t+\alpha)$.  The total pumped current $I$
into the right lead (see Fig.~\ref{fig:MZIschematic}) can then be
expressed as an integral over the area $A$ that is enclosed in
$(\tilde{\theta}_1$, $\tilde{\theta}_2)$--parameter space during one
period, and is given by the scattering matrix
expression~\cite{Blaauboer2002,Wang2002}:
\begin{subequations}
\begin{eqnarray}                                                      
  I_{p,R} & = & \frac{\omega e}{2 \pi^2} \int_A\, d\theta_1\,
  d\theta_2\, \sum_{\begin{subarray}{l} m\in L \\ n\in
      R\end{subarray}} \text{Im} \{ \Pi_{nm} (\theta_1,\theta_2)\}
  \label{eq:MZIp} \\             
  & \approx & \frac{\omega e}{2 \pi}\, \delta \theta_1\, \delta
  \theta_2\, \sin \alpha \, \sum_{\begin{subarray}{l} m\in L \\ n\in
      R\end{subarray}} \text{Im} \{ \Pi_{nm} (\theta_1,\theta_2)\}.
\label{eq:MZIpLinearResp}                                  
\end{eqnarray}                                                        
\end{subequations}                                                    
Here                                                                 
\begin{equation}                                                      
\Pi_{nm} (\theta_1,\theta_2)= \left(  \frac{\partial
  S_{\text{RL}}^{he}}{\partial \theta_1}\frac{\partial
  S_{\text{RL}}^{he*}}{\partial \theta_2} -\frac{\partial
  S_{\text{RL}}^{ee}}{\partial \theta_1}\frac{\partial
  S_{\text{RL}}^{ee*}}{\partial \theta_2} \right)_{nm}.
\label{eq:MZImIp}                                                
\end{equation}   
\eq{MZIpLinearResp} is valid in the bilinear response regime where
$\delta \theta_1 \ll\theta_{1}$ and $\delta \theta_2 \ll \theta_{2} $
and the integral in \eq{MZIp} becomes independent of the pumping
contour. $S_{\text{RL},nm}$ describes the scattering of a Dirac
fermion in mode $m$ in the left (L) lead to a Dirac fermion in mode
$n$ in the right (R) lead. The explicit adiabatic condition for this
system is given by $\hbar \omega \ll$
\{$\Delta,M,\hbar (v_m)_k/L_k$\} ($k\in\{a,b,c,d\}$).

After calculating the derivatives of the scattering matrix
coefficients using \eq{MZscatteringmatrix} and taking the imaginary
part of the product, we obtain for the pumped current into the right
lead of the single-mode pump of Fig.~\ref{fig:MZIschematic}:
\begin{equation}
  \label{eq:MZpumpedcurrent}
  I_{p,R} = I_0 \; \left( rt - 2 rt \sin^2 \left( \frac{\theta_2}{2}
  \right) - t^2 \sin(\theta_1 - \theta_2)\right),
\end{equation}
where $I_0 = (\omega e)/(4 \pi)\, \delta \theta_1\, \delta \theta_2\, \sin
\alpha$. 
From Eq.~(\ref{eq:MZpumpedcurrent}) we see that the electron (hole)
charge collected in the right lead is always mediated by the
interference of two chiral Majorana modes (see also
Ref.~\onlinecite{Li2012}).  Notice that we can rewrite the pumped
current as a sum of two terms $I_{p,R}(\theta_1,\theta_2) / I_0=
I^{(0)}(0,0)+ I^{(\theta)}(\theta_1,\theta_2)$ consisting of an
Aharonov--Bohm flux--dependent part $I^{(\theta)}(\theta_1,\theta_2)$,
and a flux--independent part $I^{(0)}(0,0)$. The latter is given by $
I^{(0)} (0,0)= r t$ which reaches its maximum value of $I^{(0)} (0,0)
= 1/2$ at $t = 1/\sqrt{2}$.
The flux--dependent part is a sum of two terms. The second term on the
right-hand side of Eq.~(\ref{eq:MZpumpedcurrent}) is proportional to
$rt$ and only depends on $\theta_2$. This is a consequence of the
chiral nature of transport: If we reverse the direction of transport,
this term will only depend on the phase $\theta_1$.
From Eq.~(\ref{eq:MZpumpedcurrent}) we also see that the QPC plays an
essential role in generating a pumped current. For a closed QPC
($t=0$) no net pumped current is generated.

For carriers in the low--energy regime, $E \ll \hbar/ \delta \tau_i$,
we can approximate $\theta_i = (n_{\nu_i} + 1) \pi$.  For a
transparent point contact, $t=1$, the pumped current then reduces to
$I_{p,R}(\theta_1,\theta_2) = 0$ for $\delta n_{-}= n_{\nu_1} -
n_{\nu_2}$ integer and it achieves maximum values
$I_{p,R}(\theta_1,\theta_2) / I_0 =~\pm~1$ for $\delta n_{-}$
half-integer.
In the latter, the pump produces a unit $I_0$ of pumped current
which is also its maximum value (for $t=1$). 
This is also true for $t = 1/\sqrt{2}$ at $\theta_1 = -\pi/2$ and
$\theta_2 = 0$ (modulo $2\pi$).
The pumped current reaches a global maximum value of
$I_{p,R} / I_0 = \pm (1 + \sqrt{2})/2 \approx 1.2$
at $t = \sqrt{2 + \sqrt{2}}/2$.

\begin{figure}
  \centering
  \includegraphics[width=.7\columnwidth]{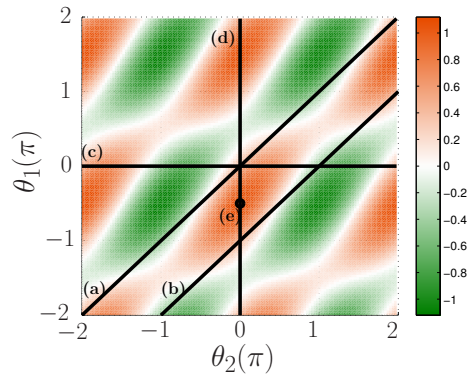}
  \caption{(color online). Contour plot of the pumped current $
    I_{p,R} /I_0$ [\eq{MZpumpedcurrent}] as a function of the phases
    $\theta_1$ and $\theta_2$ for $t = 0.8$. (a) $\theta_2=\theta_1$
    line, (b) $\theta_2 = \theta_1 - \pi$ line, (c) $\theta_1 = 0$
    line, (d) $\theta_2 = 0$ line, and (e) $\theta_1 = -\pi/2$ and
    $\theta_2 = 0$ point. The maximum (minimum) values $I_{p,R} /I_0 =
    \pm 1.12$.}
  \label{fig:Ip2Dsurf}
\end{figure}
Figure~\ref{fig:Ip2Dsurf} shows the pumped current as a function of
the phase shifts accumulated while traveling around the first and the
second superconducting islands for a fixed value of the transparency
$t$ of the QPC.
The pumped current $I_{p,R}$ clearly is a $2\pi$--periodic
function with respect to $\theta_1$ and $\theta_2$.
However, as expected from \eq{MZpumpedcurrent}, the
pumped current is not a symmetric function under exchange $\theta_1
\leftrightarrow \theta_2$.
The pumped current oscillates between positive and negatives values,
meaning that the interferometer transmits alternatively electrons and
holes.
As discussed above, in the low-energy regime the pumped current has
values near zero when the phase difference $\theta_1 - \theta_2$ is an
even or an odd multiple of $\pi$, see line~(a) and (b) in
Fig.~\ref{fig:Ip2Dsurf}.  The asymmetry between the two phases can be
seen from the difference between line~(c) and (d) in
Fig.~\ref{fig:Ip2Dsurf}.  The pumped current exhibits maximum values
at $\theta_1 = -\pi/2$ and $\theta_2 = 0$ (modulo $2\pi$), see
dot~(e).

\textit{Conductance}.\ In this section we discuss the difference
between the conductance and the pumped current in our system. This is
of importance for being able to measure the pumped current, as the
main bottleneck for its detection is the difficulty to distinguish
between the two types of currents.
Using the Landauer--B\"uttiker formalism~\cite{Nazarov2009} the
conductance across the interferometer can be written as:
\begin{equation}
  G(eV) =\frac{e^2}{h}  \left(| S_{RL}^{ee}|^2 - | S_{RL}^{he}|^2 \right).
\end{equation}
Using the scattering matrix~\eq{MZscatteringmatrix}, the conductance
is then given by:
\begin{equation}
G(eV)=  \frac{e^2}{h}\left( 1 - 2 \left[ t^2 \sin^2 \left( \frac{
      \delta \theta_-}{2} \right) + r^2 \sin^2\left( \frac{ \delta
      \theta_+}{2}\right) \right] \right),
\label{eq:MZIconductance}
\end{equation}
where $\delta \theta_\pm \equiv \theta_1 \pm \theta_2 $.  In the
low--energy regime, $E\ll\hbar /\delta \tau_i$, the conductance
reaches the limiting values:

 \begin{center}
   \begin{tabular}{c|c|c}

            $G(0)$ & $t = 0$ ($\delta n_+$)& $t = 1$ ($\delta n_-$)\\
            \hline

            $\delta n_{\pm}$ even & $+e^2/h$
            & $+e^2/h$
            \\
            \hline
            $\delta n_{\pm}$ odd & $-e^2/h$
            & $-e^2/h$
            \\
   \end{tabular}
 \end{center}
where $\delta n_\pm \equiv n_{\nu_1} \pm n_{\nu_2}$. 
When $\delta n_\pm$ is an even number the Majorana states traveling
along the two paths are unperturbed and the right normal lead collects
an electron.
When $\delta n_\pm$ is an odd number, one of the Majorana modes has
acquired an additional phase of $\pi$ and the right lead collects a
hole due to crossed Andreev reflection in which a 2$e$ charge is
absorbed by the superconductors.
In both situations, the conductance is sensitive to the number of
vortices encircled in the interferometer. This is in agreement with
the single Mach--Zehnder interferometer studied
earlier~\cite{Akhmerov2009, Fu2009}.

\begin{figure}
  \centering
  \includegraphics[width=.7\columnwidth]{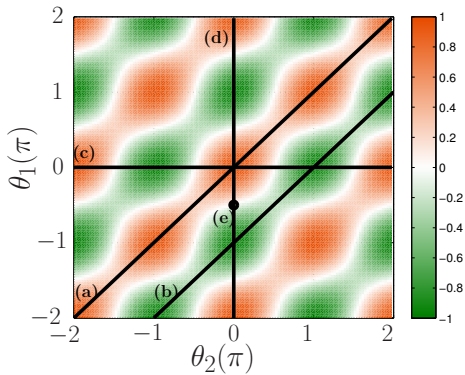}
  \caption{(color online). Contour plot of the conductance $G$
    [\eq{MZIconductance}] in units of $e^2 /h$ as a function of the
    phases $\theta_1$ and $\theta_2$ for $t =0.8$. (a)
    $\theta_2=\theta_1$ line, (b) $\theta_2 = \theta_1 - \pi$ line,
    (c) $\theta_1 = 0$ line, (d) $\theta_2 = 0$ line, and (e)
    $\theta_1 = -\pi/2$ and $\theta_2 = 0$ point.}
\label{fig:G2Dsurf}
\end{figure}

As for the pumped current, the conductance has two contributions: an
Aharonov--Bohm flux--independent part and a flux--dependent part,
$G(eV) = G^{(0)} (0,0) + G^{(\theta)} (\theta_1, \theta_2)$. The
flux--independent term is $G^{(0)} (0,0) =e^2/h$, in which the
incident electron is transmitted as an electron. The flux--dependent
term has two terms which depend, resp., on the sum and difference of
the phases, $\delta \theta_+$ and $\delta \theta_-$.  If we reverse
the direction of transport, the conductance has the same dependence on
$\theta_i$ for transport from left to right and from right to
left. Thus, the conductance reveals no signature of the chiral nature
of transport.

Figure~\ref{fig:G2Dsurf} shows the conductance in units of $e^2/h$ as
a function of the phase accumulated in the first and second
superconducting islands for a fixed value of the transparency of the
point contact.
Like the pumped current, the conductance is a $2\pi$--periodic
function with respect to $\theta_1$ and $\theta_2$. 
An interesting situation to analyze is when the interferometer does
not transmit any charge, i.e., $G(eV) = 0$. 
This happens in two different situations: first, when the point
contact is completely transparent (reflective), $t=1$($0$), and
$\delta \theta_\pm$ is a half-integer of $\pi$; and second, when $t =
r = 1/\sqrt{2}$, $\theta_1 = -\pi/2$ and $\theta_2 = 0$ (modulo
$2\pi$). 
In both cases, the processes of transmitting an electron and
transmitting a hole have the same probability to occur. Since these
two processes have opposite charge contributions, on average the total
charge collected in the right lead is zero. Interestingly, in the
second case, at these same points in parameter space the pump
generates a maximum current, as discussed earlier.

We thus predict three main differences between the conductance and the
pumped current in this system: 1) Although the conductance and the
pumped current both contain a flux--independent part and a
flux--dependent part, the flux--independent part of the conductance is
independent of any system parameters while the flux--independent part
of the pumped current depends on the transparency of the point
contact.
2) While the conductance is insensitive to the direction of transport,
the magnitude of the pumped current depends on whether the current is
collected in the right or left lead, thereby reflecting the chiral
nature of the transport.
3) At certain points in parameter space (i.e. for certain values of
$t$, $\theta_1$ and $\theta_2$), the conductance is zero, whereas the
pumped current reaches maximum values.

The proposed pumping mechanism requires the phases $\theta_1$ and
$\theta_2$ to be varied in a periodic manner.
One way to achieve this would be by periodically varying the magnetic
field in each superconducting island.
Alternatively, the velocity of the chiral Majorana states could be
changed, using the method proposed in Ref.~\onlinecite{Tiwari2012}.

\textit{Conclusions}.\ To summarize, we have analyzed quantum pumping
via Majorana fermions in a Mach-Zehnder interferometer formed by
ferromagnetic and superconducting regions on top of a 3D topological
insulator.
We have shown that in the low-energy regime the pumped current, unlike
the conductance, cannot be used to distinguish between an even or odd
number of Majorana bound states at the vortex cores in the
superconducting islands.
The pumped current, however, can be used to reveal signatures of the
chiral nature of transport, whereas the conductance is independent of
the direction of transport.
We have also shown that the pumped current reaches maximum values 
in certain regions of parameter space where the conductance
becomes zero.  Tuning the system into the latter regions thus creates
chances for experimentally observing the adiabatically pumped current
induced by Majorana modes in this system.

\textit{Acknowledgments}.\ We would like to thank A. Saha and
C. Bruder for valuable discussions.  This research was supported by
the Dutch Science Foundation NWO/FOM. RPT acknowledges financial
support by the Swiss SNF and the NCCR Quantum Science and Technology.


\begin{thebibliography}{33}
\expandafter\ifx\csname natexlab\endcsname\relax\def\natexlab#1{#1}\fi
\expandafter\ifx\csname bibnamefont\endcsname\relax
  \def\bibnamefont#1{#1}\fi
\expandafter\ifx\csname bibfnamefont\endcsname\relax
  \def\bibfnamefont#1{#1}\fi
\expandafter\ifx\csname citenamefont\endcsname\relax
  \def\citenamefont#1{#1}\fi
\expandafter\ifx\csname url\endcsname\relax
  \def\url#1{\texttt{#1}}\fi
\expandafter\ifx\csname urlprefix\endcsname\relax\def\urlprefix{URL }\fi
\providecommand{\bibinfo}[2]{#2}
\providecommand{\eprint}[2][]{\url{#2}}

\bibitem[{\citenamefont{Alicea}(2012)}]{Alicea2012}
\bibinfo{author}{\bibfnamefont{J.}~\bibnamefont{Alicea}},
  \bibinfo{journal}{Reports on Progress in Physics}
  \textbf{\bibinfo{volume}{75}}, \bibinfo{pages}{076501}
  (\bibinfo{year}{2012}).

\bibitem[{\citenamefont{Leijnse and Flensberg}(2012)}]{Leijnse2012}
\bibinfo{author}{\bibfnamefont{M.}~\bibnamefont{Leijnse}} \bibnamefont{and}
  \bibinfo{author}{\bibfnamefont{K.}~\bibnamefont{Flensberg}},
  \bibinfo{journal}{Semicond. Sci. Technol.} \textbf{\bibinfo{volume}{27}},
  \bibinfo{pages}{124003} (\bibinfo{year}{2012}).

\bibitem[{\citenamefont{Beenakker}(2013)}]{Beenakker2011}
\bibinfo{author}{\bibfnamefont{C.}~\bibnamefont{Beenakker}},
  \bibinfo{journal}{Annual Review of Condensed Matter Physics}
  \textbf{\bibinfo{volume}{4}}, \bibinfo{pages}{113} (\bibinfo{year}{2013}).

\bibitem[{\citenamefont{Moore and Read}(1991)}]{Moore1991}
\bibinfo{author}{\bibfnamefont{G.}~\bibnamefont{Moore}} \bibnamefont{and}
  \bibinfo{author}{\bibfnamefont{N.}~\bibnamefont{Read}},
  \bibinfo{journal}{Nucl. Phys.} \textbf{\bibinfo{volume}{360}},
  \bibinfo{pages}{362} (\bibinfo{year}{1991}).

\bibitem[{\citenamefont{{Stern}}(2010)}]{Stern2001}
\bibinfo{author}{\bibfnamefont{A.}~\bibnamefont{{Stern}}},
  \bibinfo{journal}{\nat} \textbf{\bibinfo{volume}{464}}, \bibinfo{pages}{187}
  (\bibinfo{year}{2010}).

\bibitem[{\citenamefont{Ivanov}(2001)}]{Ivanov2001}
\bibinfo{author}{\bibfnamefont{D.~A.} \bibnamefont{Ivanov}},
  \bibinfo{journal}{Phys. Rev. Lett.} \textbf{\bibinfo{volume}{86}},
  \bibinfo{pages}{268} (\bibinfo{year}{2001}).

\bibitem[{\citenamefont{Sau et~al.}(2010{\natexlab{a}})\citenamefont{Sau,
  Lutchyn, Tewari, and Das~Sarma}}]{Sau2010a}
\bibinfo{author}{\bibfnamefont{J.~D.} \bibnamefont{Sau}},
  \bibinfo{author}{\bibfnamefont{R.~M.} \bibnamefont{Lutchyn}},
  \bibinfo{author}{\bibfnamefont{S.}~\bibnamefont{Tewari}}, \bibnamefont{and}
  \bibinfo{author}{\bibfnamefont{S.}~\bibnamefont{Das~Sarma}},
  \bibinfo{journal}{Phys. Rev. Lett.} \textbf{\bibinfo{volume}{104}},
  \bibinfo{pages}{040502} (\bibinfo{year}{2010}{\natexlab{a}}).

\bibitem[{\citenamefont{Sau et~al.}(2010{\natexlab{b}})\citenamefont{Sau,
  Tewari, Lutchyn, Stanescu, and Das~Sarma}}]{Sau2010b}
\bibinfo{author}{\bibfnamefont{J.~D.} \bibnamefont{Sau}},
  \bibinfo{author}{\bibfnamefont{S.}~\bibnamefont{Tewari}},
  \bibinfo{author}{\bibfnamefont{R.~M.} \bibnamefont{Lutchyn}},
  \bibinfo{author}{\bibfnamefont{T.~D.} \bibnamefont{Stanescu}},
  \bibnamefont{and}
  \bibinfo{author}{\bibfnamefont{S.}~\bibnamefont{Das~Sarma}},
  \bibinfo{journal}{Phys. Rev. B} \textbf{\bibinfo{volume}{82}},
  \bibinfo{pages}{214509} (\bibinfo{year}{2010}{\natexlab{b}}).

\bibitem[{\citenamefont{Alicea}(2010)}]{Alicea2010}
\bibinfo{author}{\bibfnamefont{J.}~\bibnamefont{Alicea}},
  \bibinfo{journal}{Phys. Rev. B} \textbf{\bibinfo{volume}{81}},
  \bibinfo{pages}{125318} (\bibinfo{year}{2010}).

\bibitem[{\citenamefont{Hasan and Kane}(2010)}]{Hasan2010}
\bibinfo{author}{\bibfnamefont{M.~Z.} \bibnamefont{Hasan}} \bibnamefont{and}
  \bibinfo{author}{\bibfnamefont{C.~L.} \bibnamefont{Kane}},
  \bibinfo{journal}{Rev. Mod. Phys.} \textbf{\bibinfo{volume}{82}},
  \bibinfo{pages}{3045} (\bibinfo{year}{2010}).

\bibitem[{\citenamefont{Qi and Zhang}(2011)}]{Qi2011}
\bibinfo{author}{\bibfnamefont{X.-L.} \bibnamefont{Qi}} \bibnamefont{and}
  \bibinfo{author}{\bibfnamefont{S.-C.} \bibnamefont{Zhang}},
  \bibinfo{journal}{Rev. Mod. Phys.} \textbf{\bibinfo{volume}{83}},
  \bibinfo{pages}{1057} (\bibinfo{year}{2011}).

\bibitem[{\citenamefont{Fu and Kane}(2008)}]{Fu2008}
\bibinfo{author}{\bibfnamefont{L.}~\bibnamefont{Fu}} \bibnamefont{and}
  \bibinfo{author}{\bibfnamefont{C.~L.} \bibnamefont{Kane}},
  \bibinfo{journal}{Phys. Rev. Lett.} \textbf{\bibinfo{volume}{100}},
  \bibinfo{pages}{096407} (\bibinfo{year}{2008}).

\bibitem[{\citenamefont{Tiwari et~al.}(2013)\citenamefont{Tiwari, Z\"ulicke,
  and Bruder}}]{Tiwari2012}
\bibinfo{author}{\bibfnamefont{R.~P.} \bibnamefont{Tiwari}},
  \bibinfo{author}{\bibfnamefont{U.}~\bibnamefont{Z\"ulicke}},
  \bibnamefont{and} \bibinfo{author}{\bibfnamefont{C.}~\bibnamefont{Bruder}},
  \bibinfo{journal}{Phys. Rev. Lett.} \textbf{\bibinfo{volume}{110}},
  \bibinfo{pages}{186805} (\bibinfo{year}{2013}).

\bibitem[{\citenamefont{Sengupta et~al.}(2001)\citenamefont{Sengupta, \ifmmode
  \check{Z}\else \v{Z}\fi{}uti\ifmmode~\acute{c}\else \'{c}\fi{}, Kwon,
  Yakovenko, and Das~Sarma}}]{zba1}
\bibinfo{author}{\bibfnamefont{K.}~\bibnamefont{Sengupta}},
  \bibinfo{author}{\bibfnamefont{I.}~\bibnamefont{\ifmmode \check{Z}\else
  \v{Z}\fi{}uti\ifmmode~\acute{c}\else \'{c}\fi{}}},
  \bibinfo{author}{\bibfnamefont{H.-J.} \bibnamefont{Kwon}},
  \bibinfo{author}{\bibfnamefont{V.~M.} \bibnamefont{Yakovenko}},
  \bibnamefont{and}
  \bibinfo{author}{\bibfnamefont{S.}~\bibnamefont{Das~Sarma}},
  \bibinfo{journal}{Phys. Rev. B} \textbf{\bibinfo{volume}{63}},
  \bibinfo{pages}{144531} (\bibinfo{year}{2001}).

\bibitem[{\citenamefont{Bolech and Demler}(2007)}]{zba2}
\bibinfo{author}{\bibfnamefont{C.~J.} \bibnamefont{Bolech}} \bibnamefont{and}
  \bibinfo{author}{\bibfnamefont{E.}~\bibnamefont{Demler}},
  \bibinfo{journal}{Phys. Rev. Lett.} \textbf{\bibinfo{volume}{98}},
  \bibinfo{pages}{237002} (\bibinfo{year}{2007}).

\bibitem[{\citenamefont{Law et~al.}(2009)\citenamefont{Law, Lee, and
  Ng}}]{Law2009}
\bibinfo{author}{\bibfnamefont{K.~T.} \bibnamefont{Law}},
  \bibinfo{author}{\bibfnamefont{P.~A.} \bibnamefont{Lee}}, \bibnamefont{and}
  \bibinfo{author}{\bibfnamefont{T.~K.} \bibnamefont{Ng}},
  \bibinfo{journal}{Phys. Rev. Lett.} \textbf{\bibinfo{volume}{103}},
  \bibinfo{pages}{237001} (\bibinfo{year}{2009}).

\bibitem[{\citenamefont{Flensberg}(2010)}]{zba3}
\bibinfo{author}{\bibfnamefont{K.}~\bibnamefont{Flensberg}},
  \bibinfo{journal}{Phys. Rev. B} \textbf{\bibinfo{volume}{82}},
  \bibinfo{pages}{180516} (\bibinfo{year}{2010}).

\bibitem[{\citenamefont{Mourik et~al.}(2012)\citenamefont{Mourik, Zuo, Frolov,
  Plissard, Bakkers, and Kouwenhoven}}]{Mourik2012}
\bibinfo{author}{\bibfnamefont{V.}~\bibnamefont{Mourik}},
  \bibinfo{author}{\bibfnamefont{K.}~\bibnamefont{Zuo}},
  \bibinfo{author}{\bibfnamefont{S.~M.} \bibnamefont{Frolov}},
  \bibinfo{author}{\bibfnamefont{S.~R.} \bibnamefont{Plissard}},
  \bibinfo{author}{\bibfnamefont{E.~. P. A.~M.} \bibnamefont{Bakkers}},
  \bibnamefont{and} \bibinfo{author}{\bibfnamefont{L.~P.}
  \bibnamefont{Kouwenhoven}}, \bibinfo{journal}{Science}
  \textbf{\bibinfo{volume}{336}}, \bibinfo{pages}{1003} (\bibinfo{year}{2012}).

\bibitem[{\citenamefont{Das et~al.}(2012)\citenamefont{Das, Ronen, Most, Oreg,
  Heiblum, and Shtrikman}}]{Das2012}
\bibinfo{author}{\bibfnamefont{A.}~\bibnamefont{Das}},
  \bibinfo{author}{\bibfnamefont{Y.}~\bibnamefont{Ronen}},
  \bibinfo{author}{\bibfnamefont{Y.}~\bibnamefont{Most}},
  \bibinfo{author}{\bibfnamefont{Y.}~\bibnamefont{Oreg}},
  \bibinfo{author}{\bibfnamefont{M.}~\bibnamefont{Heiblum}}, \bibnamefont{and}
  \bibinfo{author}{\bibfnamefont{H.}~\bibnamefont{Shtrikman}},
  \bibinfo{journal}{Nature Physics} \textbf{\bibinfo{volume}{8}},
  \bibinfo{pages}{887 –} (\bibinfo{year}{2012}).

\bibitem[{\citenamefont{Deng et~al.}(2012)\citenamefont{Deng, Yu, Huang,
  Larsson, Caroff, and Xu}}]{Deng2012}
\bibinfo{author}{\bibfnamefont{M.~T.} \bibnamefont{Deng}},
  \bibinfo{author}{\bibfnamefont{C.~L.} \bibnamefont{Yu}},
  \bibinfo{author}{\bibfnamefont{G.~Y.} \bibnamefont{Huang}},
  \bibinfo{author}{\bibfnamefont{M.}~\bibnamefont{Larsson}},
  \bibinfo{author}{\bibfnamefont{P.}~\bibnamefont{Caroff}}, \bibnamefont{and}
  \bibinfo{author}{\bibfnamefont{H.~Q.} \bibnamefont{Xu}},
  \bibinfo{journal}{Nano Lett.} \textbf{\bibinfo{volume}{12}},
  \bibinfo{pages}{6414} (\bibinfo{year}{2012}).

\bibitem[{\citenamefont{Rokhinson et~al.}(2012)\citenamefont{Rokhinson, Liu,
  and Furdyna}}]{Rokhinson2012}
\bibinfo{author}{\bibfnamefont{L.~P.} \bibnamefont{Rokhinson}},
  \bibinfo{author}{\bibfnamefont{X.}~\bibnamefont{Liu}}, \bibnamefont{and}
  \bibinfo{author}{\bibfnamefont{J.~K.} \bibnamefont{Furdyna}},
  \bibinfo{journal}{Nat. Phys.} \textbf{\bibinfo{volume}{8}},
  \bibinfo{pages}{795} (\bibinfo{year}{2012}).

\bibitem[{\citenamefont{Williams et~al.}(2012)\citenamefont{Williams, Bestwick,
  Gallagher, Hong, Cui, Bleich, Analytis, Fisher, and
  Goldhaber-Gordon}}]{Williams2012}
\bibinfo{author}{\bibfnamefont{J.~R.} \bibnamefont{Williams}},
  \bibinfo{author}{\bibfnamefont{A.~J.} \bibnamefont{Bestwick}},
  \bibinfo{author}{\bibfnamefont{P.}~\bibnamefont{Gallagher}},
  \bibinfo{author}{\bibfnamefont{S.~S.} \bibnamefont{Hong}},
  \bibinfo{author}{\bibfnamefont{Y.}~\bibnamefont{Cui}},
  \bibinfo{author}{\bibfnamefont{A.~S.} \bibnamefont{Bleich}},
  \bibinfo{author}{\bibfnamefont{J.~G.} \bibnamefont{Analytis}},
  \bibinfo{author}{\bibfnamefont{I.~R.} \bibnamefont{Fisher}},
  \bibnamefont{and}
  \bibinfo{author}{\bibfnamefont{D.}~\bibnamefont{Goldhaber-Gordon}},
  \bibinfo{journal}{Phys. Rev. Lett.} \textbf{\bibinfo{volume}{109}},
  \bibinfo{pages}{056803} (\bibinfo{year}{2012}).

\bibitem[{\citenamefont{Fu and Kane}(2009)}]{Fu2009}
\bibinfo{author}{\bibfnamefont{L.}~\bibnamefont{Fu}} \bibnamefont{and}
  \bibinfo{author}{\bibfnamefont{C.~L.} \bibnamefont{Kane}},
  \bibinfo{journal}{Phys. Rev. Lett.} \textbf{\bibinfo{volume}{102}},
  \bibinfo{pages}{216403} (\bibinfo{year}{2009}).

\bibitem[{\citenamefont{Akhmerov et~al.}(2009)\citenamefont{Akhmerov, Nilsson,
  and Beenakker}}]{Akhmerov2009}
\bibinfo{author}{\bibfnamefont{A.~R.} \bibnamefont{Akhmerov}},
  \bibinfo{author}{\bibfnamefont{J.}~\bibnamefont{Nilsson}}, \bibnamefont{and}
  \bibinfo{author}{\bibfnamefont{C.~W.~J.} \bibnamefont{Beenakker}},
  \bibinfo{journal}{Phys. Rev. Lett.} \textbf{\bibinfo{volume}{102}},
  \bibinfo{pages}{216404} (\bibinfo{year}{2009}).

\bibitem[{\citenamefont{Str\"ubi et~al.}(2011)\citenamefont{Str\"ubi, Belzig,
  Choi, and Bruder}}]{Strubi2011}
\bibinfo{author}{\bibfnamefont{G.}~\bibnamefont{Str\"ubi}},
  \bibinfo{author}{\bibfnamefont{W.}~\bibnamefont{Belzig}},
  \bibinfo{author}{\bibfnamefont{M.-S.} \bibnamefont{Choi}}, \bibnamefont{and}
  \bibinfo{author}{\bibfnamefont{C.}~\bibnamefont{Bruder}},
  \bibinfo{journal}{Phys. Rev. Lett.} \textbf{\bibinfo{volume}{107}},
  \bibinfo{pages}{136403} (\bibinfo{year}{2011}).

\bibitem[{\citenamefont{B{\"u}ttiker et~al.}(1994)\citenamefont{B{\"u}ttiker,
  Thomas, and Pr{\^e}tre}}]{Buttiker1994}
\bibinfo{author}{\bibfnamefont{M.}~\bibnamefont{B{\"u}ttiker}},
  \bibinfo{author}{\bibfnamefont{H.}~\bibnamefont{Thomas}}, \bibnamefont{and}
  \bibinfo{author}{\bibfnamefont{A.}~\bibnamefont{Pr{\^e}tre}},
  \bibinfo{journal}{Z. Phys. B} \textbf{\bibinfo{volume}{94}},
  \bibinfo{pages}{133} (\bibinfo{year}{1994}).

\bibitem[{\citenamefont{Brouwer}(1998)}]{Brouwer1998}
\bibinfo{author}{\bibfnamefont{P.~W.} \bibnamefont{Brouwer}},
  \bibinfo{journal}{Phys. Rev. B} \textbf{\bibinfo{volume}{58}},
  \bibinfo{pages}{10135} (\bibinfo{year}{1998}).

\bibitem[{\citenamefont{Spivak et~al.}(1995)\citenamefont{Spivak, Zhou, and
  Monod}}]{Spivak1995}
\bibinfo{author}{\bibfnamefont{B.}~\bibnamefont{Spivak}},
  \bibinfo{author}{\bibfnamefont{F.}~\bibnamefont{Zhou}}, \bibnamefont{and}
  \bibinfo{author}{\bibfnamefont{M.~T.~B.} \bibnamefont{Monod}},
  \bibinfo{journal}{Phys. Rev. B} \textbf{\bibinfo{volume}{51}},
  \bibinfo{pages}{13226} (\bibinfo{year}{1995}).

\bibitem[{\citenamefont{{Gibertini} et~al.}(2013)\citenamefont{{Gibertini},
  {Fazio}, {Polini}, and {Taddei}}}]{Gibertini2013}
\bibinfo{author}{\bibfnamefont{M.}~\bibnamefont{{Gibertini}}},
  \bibinfo{author}{\bibfnamefont{R.}~\bibnamefont{{Fazio}}},
  \bibinfo{author}{\bibfnamefont{M.}~\bibnamefont{{Polini}}}, \bibnamefont{and}
  \bibinfo{author}{\bibfnamefont{F.}~\bibnamefont{{Taddei}}},
  \bibinfo{journal}{ArXiv e-prints}  (\bibinfo{year}{2013}),
  \eprint{1302.2736}.

\bibitem[{\citenamefont{Blaauboer}(2002)}]{Blaauboer2002}
\bibinfo{author}{\bibfnamefont{M.}~\bibnamefont{Blaauboer}},
  \bibinfo{journal}{Phys. Rev. B} \textbf{\bibinfo{volume}{65}},
  \bibinfo{pages}{235318} (\bibinfo{year}{2002}).

\bibitem[{\citenamefont{Wang and Wang}(2002)}]{Wang2002}
\bibinfo{author}{\bibfnamefont{J.}~\bibnamefont{Wang}} \bibnamefont{and}
  \bibinfo{author}{\bibfnamefont{B.}~\bibnamefont{Wang}},
  \bibinfo{journal}{Phys. Rev. B} \textbf{\bibinfo{volume}{65}},
  \bibinfo{pages}{153311} (\bibinfo{year}{2002}).

\bibitem[{\citenamefont{Li et~al.}(2012)\citenamefont{Li, Fleury, and
  B\"uttiker}}]{Li2012}
\bibinfo{author}{\bibfnamefont{J.}~\bibnamefont{Li}},
  \bibinfo{author}{\bibfnamefont{G.}~\bibnamefont{Fleury}}, \bibnamefont{and}
  \bibinfo{author}{\bibfnamefont{M.}~\bibnamefont{B\"uttiker}},
  \bibinfo{journal}{Phys. Rev. B} \textbf{\bibinfo{volume}{85}},
  \bibinfo{pages}{125440} (\bibinfo{year}{2012}).

\bibitem[{\citenamefont{Nazarov and Blanter}(2009)}]{Nazarov2009}
\bibinfo{author}{\bibfnamefont{Y.~V.} \bibnamefont{Nazarov}} \bibnamefont{and}
  \bibinfo{author}{\bibfnamefont{Y.~M.} \bibnamefont{Blanter}},
  \emph{\bibinfo{title}{Quantum Transport: Introduction to Nanoscience}}
  (\bibinfo{publisher}{Cambridge University Press}, \bibinfo{year}{2009}).

\end{thebibliography}
\end{document}